# Modeling and Analysis using Hybrid Petri Nets

Latefa GHOMRI

Laboratoire d'Automatique de Tlemcen, FSI
UABB, BP 230, Tlemcen, 13000,
ALGERIA.

ghomrilatefa@yahoo.fr

Hassane ALLA

Laboratoire d'Automatique de Grenoble.
INPG / CNRS ENSIEG - BP46, St Martin
d'Hères – Cedex FRANCE

hassane.alla@inpg.fr

**Abstract**

This paper is devoted to the use of Hybrid Petri nets (PNs) for modeling and control of hybrid dynamic systems (HDS). Modeling, analysis and control of HDS attract ever more researchers' attention and several works has been devoted to these topics. We consider in this paper the extensions of the PN formalism (initially conceived for modeling and analysis of discrete event systems) in the direction of hybrid modeling. We present, first, the continuous PN models. These models are obtained from discrete PNs by the fluidification of the markings. They constitute the first steps in the extension of PNs toward hybrid modeling. Then, we present two hybrid PNs models, which differ in the class of HDS they can deal with. The first one is used for deterministic HDS modeling, whereas the second one can deal with HDS with nondeterministic behavior.

**Key – words**

Hybrid dynamic systems; D – elementary hybrid Petri Nets; hybrid automata; controller synthesis.

## I.     Introduction

Hybrid dynamic systems (HDS) are dynamic systems integrating explicitly and simultaneously continuous systems and discrete event systems, which require for their description, the use of continuous time model, discrete event model and the interface between them [1]. The hybrid character of the system either can owe to the system itself or to the control applied to this system.

Branicky, Borkar and Mitters [2] identified the principal physical phenomena that can be present in a HDS and specify that they can be either autonomous or controlled. They defined: (1) V*ector field switch* that is observed when the vector field changes in a discontinuous way, when the continuous state reaches a certain level. (2) *State jump* is a discontinuous jump in the continuous state when it reaches certain area in the state space.

Modeling, analysis and control of HDS are currently attracting much attention and several works were devoted to these topics; which were tackled from two different angles. On the one hand, tools conceived for modeling and analysis of continuous systems were adapted to be able to deal with switched systems. This approach consists to integrate the event aspect within a continuous formalism. Introducing commutation elements in the Bond-graph formalism is an example of this approach. On the other hand, discrete event systems tools were extended for the modeling and analysis of HDS. In this approach, a continuous aspect is integrating in discrete event formalism. An example of such formalism is the hybrid automaton (HA); which are finite state automata, extended with real value variables evolving according to differential equations. Hybrid automata are the most general HDS formalism, since they can model the largest variety of HDS. They present the advantage of combining the basic model of continuous systems, which are differential equations, with the basic model of discrete event systems, which are finite state automata. They have, in addition, a clear graphical representation; indeed, the discrete and continuous parts are well identified. The existence of automatic tools for HA reachability analysis, such as HyTech, CMC, UPPAAL and KRONOS, confer on this formalism a great analysis power. Most verification and controller synthesis techniques use HA as the investigation tool.

In this work, we consider the extensions of PN formalism, initially a model for discrete event systems, so that it can be used for modeling and control of HDS. This is an event driven system point of view. The studied systems correspond to discrete event behaviors with simple continuous dynamics.

PNs were introduced, and are still used, for discrete event systems description and analysis [3]. Currently, much effort is devoted to adapt this formalism so that it can deal with HDS and many hybrid PNs formalisms were conceived.

The first steps in this direction were taken in [4] by introducing the first continuous PN model. Continuous PNs can be used either to describe continuous flow systems, or to provide a continuous approximation of discrete event systems behavior, in order to reduce the computing time. The marking is no more given as a vector of integers, but as a real numbers vector. Thus, during a transition firing, an infinitesimal quantity of marking is taken from upstream places and put in the downstream places. This involves that transitions firing is no more an instantaneous operation but a continuous process characterized by a speed. This speed can be compared to a flow rate. All continuous PN models defined in the literature differ only in the manner of calculating instantaneous firing speeds of transitions.

From continuous PNs, the Hybrid PN formalism was defined in [5], and since it is the first hybrid formalism to be defined from PNs, the authors, simply, gave it the name of hybrid PN. This formalism combines in the same model a continuous PN, which represents the continuous flow, and a discrete T – timed PN [6], to represent the discrete behavior.

Demongodin, Aubry and Prunet [7] extended HPN in order to introduce delays and accumulations, which allow modelling of batches characteristics and transformations. This is carried out by attribution to certain places (called batch-places) of parameters and laws of the system evolution throughout marking. In addition, speeds of transitions called batches- transition are function of information related to upstream and downstream batch-places. This type of formalisms is well adapted for modeling of manufacturing lines where there are conveyors.

The first idea of D – elementary hybrid PN, was introduced in [8] and differs from traditional hybrid PN model in the fact that it integrates a T – time discrete PN [9][10] for describing the discrete part. It confers on D – elementary hybrid PNs a nondeterministic behavior, so the model can be used for modeling open loop systems (systems not coupled to their controllers).

Differential PNs were defined in [11]. This formalism integrates two types of places and transitions: discrete and differential. This allows defining of real marking (positive, negative or null). Firing speeds and temporizations are affected to differential transitions; which allows time discretizing. It is thus possible to represent a discretized model where the continuous part is of first order (the state variables correspond to the differential places marking).We thus have linear continuous variables per pieces between each discretization step.

We consider in this paper the extensions of the PN formalism in the direction of hybrid modeling. Section 2 presents the continuous PN models. These models are obtained from discrete PNs by the fluidification of the markings. They constitute the first steps in the extension of PNs toward hybrid modeling. We will see the major advantages and drawbacks of each model. Then, Section 3 presents two hybrid PNs models, which differ in the class of HDS they can deal with. The first one is used for deterministic HDS modeling, whereas the second one can deal with HDS with nondeterministic behavior. This section constitutes the main contribution of this paper. Section IV addresses briefly the general control structure based on hybrid PNs. At last, section V gives a conclusion and our future research.

## II. Continuous Petri nets

Continuous Petri net were introduced in [4] as an extension of traditional Petri net where the marking is fluid. A transition firing is a continuous process and consequently the state equation is a differential equation. A continuous PN allows, certainly, the description of positive continuous systems, but it is also used to approximate modeling of discrete events systems (DES). The main advantage of this approximation is that the number of occurring events is considerably smaller than for the corresponding discrete PN. Moreover, the analysis of a continuous PN does not require an exhaustive enumeration of the discrete state space.

An autonomous continuous PN is defined as follows:

***Definition 1 (autonomous continuous Petri Net):*** An autonomous continuous Petri net is a structure $\mathcal{PN} = (P, T, Pre, Post, M_0)$ such that:

- $P = \{P_1, P_2, \ldots, P_n\}$ is a nonempty finite set of n places ;
- $T = \{T_1, T_2, \ldots, T_m\}$ is a nonempty finite set of m transitions ;
- $Pre : P \times T \rightarrow R^+$ is the pre – incidence function that associates a positive rational weight for each arc $(T_j, P_i)$ ;
- $Post : P \times T \rightarrow R^+$ is the post – incidence function that associates a positive rational weight for each arc $(P_i, T_j)$ ;
- $M_0 : P \rightarrow R^+$ in the initial marking ;

The following notations will be considered in the sequel:

$°T_J$ is the set of input places of transition $T_J$.

$T°_J$ is the set of output places of transition $T_J$.

❏

As in a classical PN, the state of a continuous PN is given by its marking; however, the number of continuous PN reachable markings is infinite. That brought the authors in [15] to group several markings into a macro – marking. The notion of macro – marking is defined as follows:

***Definition 2 (macro – marking)*** : Let $\mathcal{PN}$ be an autonomous continuous PN and $M_k$ its marking at time k. $M_k$ may divide P (the set of places) into two subsets :

1. $P^+(M_k)$ : The set of places with positive marking ;
2. $P^0(M_k)$ : The set of places whose marking is null ;

A Macro – marking is the set of all markings which have the same subsets $P^+$ and $P^0$. A macro – marking can be characterized by a Boolean vector as follows:

$V : \quad P \rightarrow \{0, 1\}$

$$P_i \rightarrow \begin{cases} 1 & \text{si} \quad P_i \in P^+ \\ 0 & \text{si} \quad P_i \in P^0 \end{cases}$$

❏

The concept of macro – marking was defined as a tool that permits to represent in a finite way, the infinite set of states (markings) reachable by a continuous PN. The number of reachable macro– marking of an n–place continuous PN is less than or equal to $2^n$, even if the continuous PN is unbounded, since each macro marking is based on a Boolean state. A macro–marking is denoted $m^*_j$

***Example 1 :*** Consider the system of three connected tanks shown in Figure 1.a. Tank 1 and tank 2 are supplied by valves 1 and 2. Tank 1 (tank 2) and tank 3 are connected by means of valve 3 (valve 4). It is supposed that initially, tank 1, tank 2 and tank 3 contain 25, 10 and 5 v.u. (volume units) respectively.

The continuous PN shown in Figure 1.b. describes the behavior of the system of tanks. Places and transitions of the continuous PN are represented with double line to distinguish them from places and transitions of a discrete PN. The firing of transitions $T_1$, $T_2$, $T_3$ and $T_4$ represents material flow through valve 1, valve 2, valve 3 and valve 4 respectively. The marking of places $P_1$, $P_2$ and $P_3$ represents quantities of liquid in tank 1, tank 2 and tank 3 respectively. Figure 1.c. represents the reachability graph, it contains all macro – marking reachable by the continuous PN.

From the basic definition of autonomous continuous PNs, several researchers have defined several timed continuous PNs formalisms. Among these formalisms, we present mainly the continuous model with constant maximal speeds and we give briefly some explanations about the continuous model where the speeds depend on the marking..

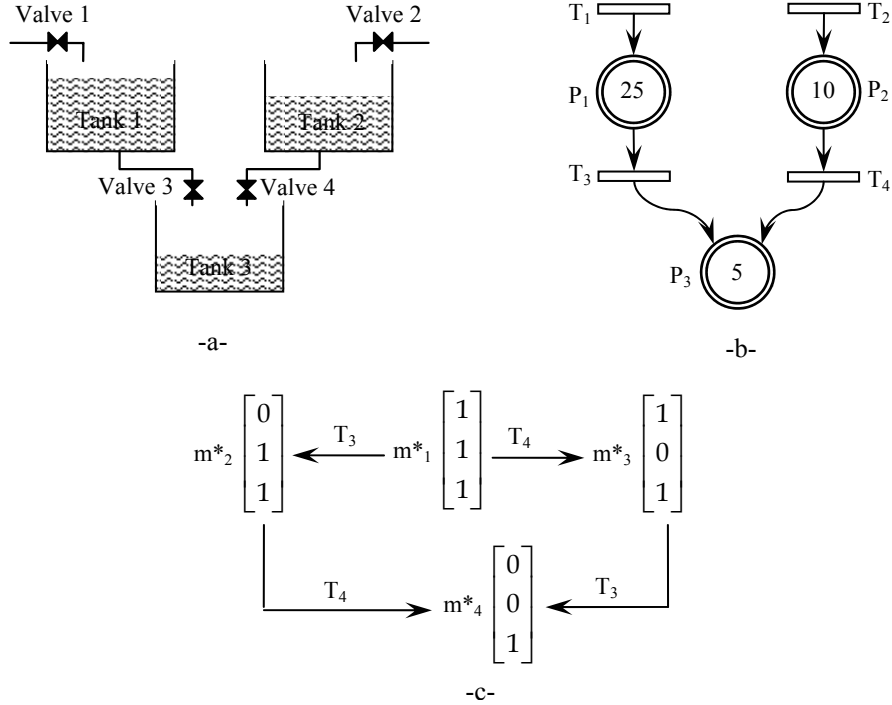

Figure 1. – a – System of tanks. – b – Continuous PN describing the system of tanks.
– c – Reachability graph for the CPN.

### II.1. Constant speed continuous Petri nets

Constant speed Continuous Petri net CCPN [15] is the first timed continuous PN to be defined. It is also the most studied model. It is defined as follows:

***Definition 3 (Constant speed continuous Petri nets)***: A constant speed continuous Petri nets is a structure $\mathcal{PN}_C = (\mathcal{PN}, V)$ such that:

- $\mathcal{PN}$ is an autonomous continuous PN.
- $V: \quad T \to R^+$

    $T_j \to V_j$

    is a function that associates to each transition $T_j$ its maximal firing speed $V_j$.

❏

In a CCPN, a place marking is a real number that evolves according to transitions instantaneous firing speeds.

An instantaneous firing speed $v_j(t)$ of a continuous transition $T_j$ can be seen as the flow of markings that crosses this transition. It lies between 0 and $V_j$ for a transition $T_j$ of a CCPN. The concept of validation of a continuous transition is different from the traditional concept met in discrete PNs. We consider that a transition of a CCPN can have two states:

1. The state strongly enabled, if

$\forall P_i \in °T_j, P_i \in P^+$

Here, the transition $T_j$ is fired at its maximal firing speed $V_j$;

2. The state weakly enabled, if

$\exists P_i \in °T_j, P_i \in P^0$

In this case, the transition $T_j$ is fired at a speed $v_j$ lower than its maximum firing speed.

The state equation in a CCPN is as follows:

$$\dot{m} = W.v(t)$$

Where $W$ is the PN incidence matrix. This implies that the evolution in time of the state of a CCPN is given by the resolution of the differential equation above, knowing the instantaneous firing speeds vector. The evolution of a CCPN in time is given by a graph whose nodes represent instantaneous firing speeds vector. Each node is called a phase. In addition, each transition is labeled with the event indicating the place, which marking becomes nil and causes the changing of speed state. The duration of a phase is also indicated. For more details, see [15]

***Example 2:*** Consider the system of three tanks (Figure 2.a.), where we associate to valve 1, valve 2, valve 3 and valve 4, the flow rates 2 v.u./t.u. (volume units by time unit) 5 v.u./t.u., 3 v.u./t.u. and 6 v.u./t.u. respectively. This system is described with the CCPN in Figure 2.b. The only difference between this model and the autonomous continuous PN (Figure 1.b.) is that with each transition is associated a maximal firing speed.

Since all the places are marked, then all the instantaneous firing speeds are equal to their maximal value. The marking balance for each place is given by the input flow minus the output flow, then:

At initial time t = 0, $v_1 = 2, v_2 = 5, v_3 = 3, v_4 = 6$, then $\dot{m}_1 = -1, \dot{m}_2 = -1$, and $\dot{m}_3 = 9$.

Markings $m_1$, $m_2$ and $m_3$ evolve initially according to the following equations, respectively:

$m_1 = 25 - t, m_2 = 10 - t$, and $m_3 = 5 + 9t$.

At time t = 10 the marking $m_2$ becomes nil, which defines a new dynamics for the system, as follows:

$v_1 = 2, v_2 = 5, v_3 = 3, v_4 = 5$, then $\dot{m}_1 = -1, \dot{m}_2 = 0$, and $\dot{m}_3 = 8$.

And after time 10, $m_1 = 15 - (t - 10), m_2 = 0$, and $m_3 = 95 + 8(t - 10)$.

In the same way, at time t = 25, the marking $m_1$ becomes nil, this defines the following dynamics:

$v_1 = 2, v_2 = 5, v_3 = 2, v_4 = 5$ then $\dot{m}_1 = 0, \dot{m}_2 = 0$, and $\dot{m}_3 = 7$.

And after time 25, $m_1 = 0, m_2 = 0$, and $m_3 = 215 + 7(t - 25)$.

The curves in figures 3.a, 3.b and 3.c schematize markings $m_1$, $m_2$ and $m_3$ dynamics. These plots are made with the software SIRPHYCO [28]. This tool permits the simulation of discrete, continuous and hybrid PNs.

The evolution of this model in time can be described thanks to the evolution graph in Figure 2.c. It can be notices that the marking of place $P_3$ is unbounded while the number of nodes is finite and equal to 3.

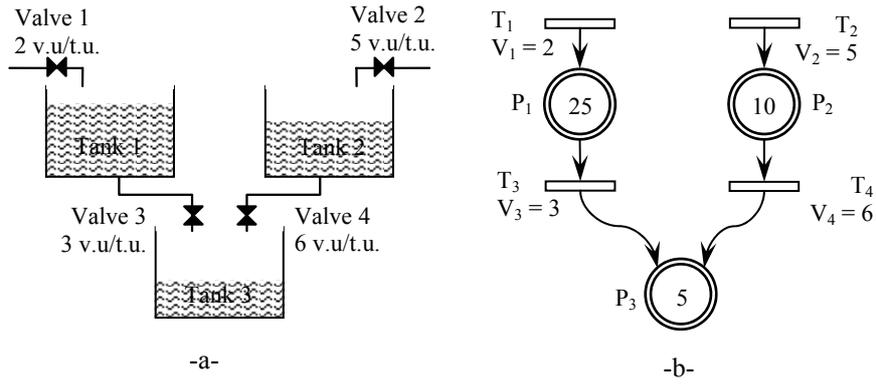

Figure 2. –a– System of tanks with valves flow rates. –b– CCPN of the system of tanks.
–c– The evolution graph

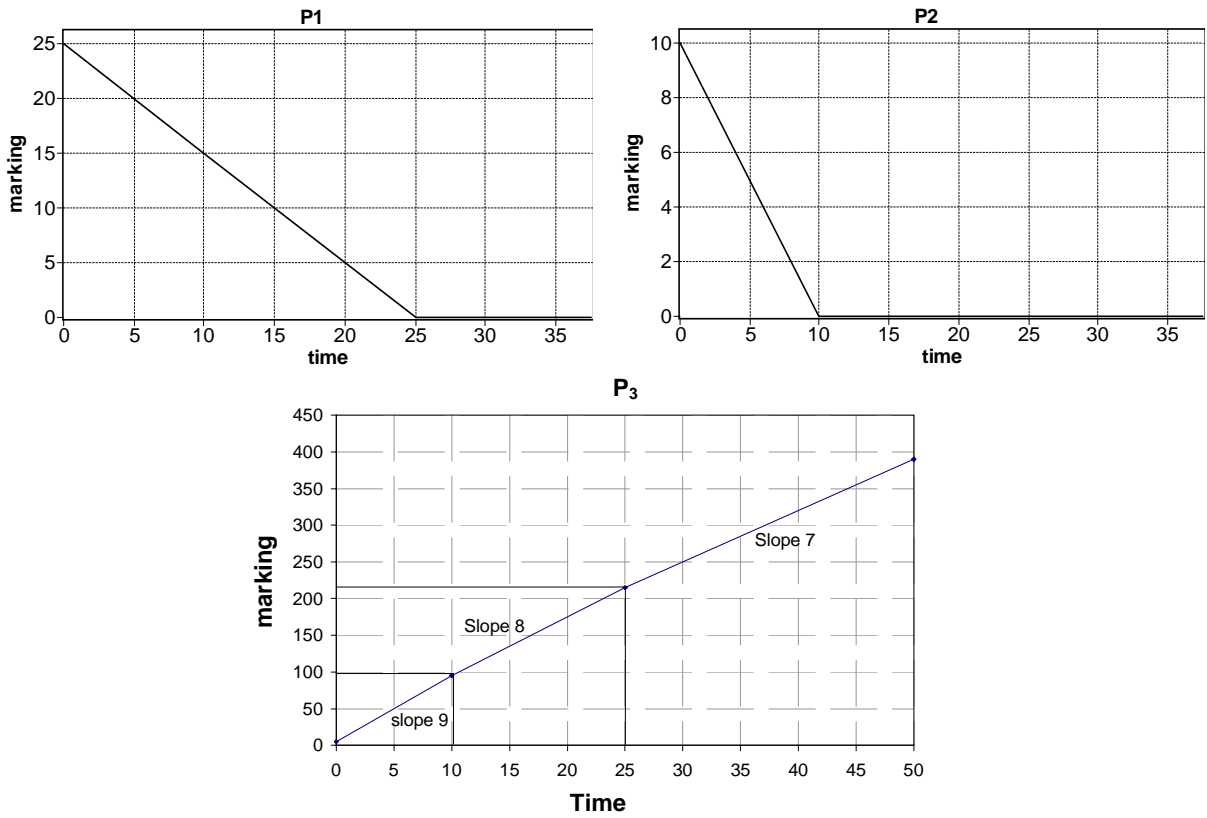

Figure 3. Temporal evolution of the marking of PN in figure 2.b.

## II.2. Variable speed continuous Petri nets

A variable speed continuous PN (VCPN) is preferred to CCPN when we want to approximate more precisely the modeling of a discrete event system [15]. It is distinguished from CCPN in the way of calculating the instantaneous firing speeds of transitions. The instantaneous firing speed of a transition Tj is obtained as follows: $v_j(t) = V_j \min_{P_i \in °T_j} m_i$

The behavior of a VCPN is given by a piecewise linear system. The passage from a linear phase to another one is driven by the function Min, which depends on the current markings. These changes correspond to autonomous commutations. The evolution of the marking of VCPN can be represented by a graph similar to that used for CCPN. The main difference is that in the evolution graph, nodes of a VCPN may contain terms of the form $\dot{m} = a.m$, where $a$ is a constant).

## II.3. Conclusion on Continuous Petri Net models

The CCPN model is used for modeling of the continuous systems where the time derivatives of the markings are constant. It approximates the modeling of discrete event systems with significant material flow. The macro–state of this model is given by the macro–markings. The only one type of events that may change the macro–state of a CCPN is that a place marking becomes nil. This event is uncontrollable.

The VCPN model was proposed to improve the approximation in modeling a discrete event system. Also in this model, only one type of events may change the macro–state when the function Min commutes, this is also an uncontrollable event.

## III. Hybrid Petri nets

Continuous PNs are used for modeling continuous flow systems, however, this model does not allow logical conditions or discrete behaviors modeling (*e.g.* a valve may be open or closed). For permitting modeling of discrete states, hybrid PNs were defined [5]. In a hybrid PN, the firing of a continuous transition describes the material flow, while the firing of a discrete transition models the occurrence of an event that can for example change firing speeds of the continuous transitions.

We find in the literature several types of continuous PNs [15] and several types of discrete PNs integrating time [6][9]. In the autonomous hybrid model definition, there are no constraints on discrete and continuous parts type. The first, and more used hybrid PN formalism to be defined, simply called hybrid Petri net combines a CCPN and a T–timed PN. The combination of these two models confers on the hybrid model a deterministic behavior. It is used for the performance evaluation of hybrid systems.

D–elementary hybrid PNs are another type of hybrid PN formalisms. They combine a time PN and a constant speed continuous PN (CCPN) [4]. Time PNs are obtained from Petri nets by associating a temporal interval with each transition. They are used as an analysis tool for time dependent systems.

However, hybrid PNs were defined before D–elementary hybrid PNs. In order to simplify the presentation, we will start by defining D–elementary hybrid PNs.

### III.1. D – elementary hybrid Petri nets

***Definition 4*** (***D-elementary hybrid PNs*****):** A D–elementary hybrid PN is a structure $\mathcal{PN_H}$ = (*P, T, Pre, Post, h, S, V, $M_0$*) such that:

1.  *P* = {$P_1, P_2, …, P_n$} is a finite set of n places;

2. $T = \{T_1, T_2, \ldots, T_m\}$ is a finite set of m transitions;

We denote $P^D = \{P_1, P_2, \ldots, P_{n'}\}$ the set of n' discrete places (denoted by D–places and drawn as simple circles) and $T^D \{T_1, T_2, \ldots, T_{m'}\}$ the set of the m' discrete transitions (denoted by D–transitions and drawn as black boxes). $P^C = P - P^D$ and $T^C = T - T^D$ denotes respectively the sets of continuous places (denoted by C–places and drawn with double circles) and continuous transitions (denoted by C–transitions and drawn as empty boxes).

3. *Pre* : P x T $\rightarrow$ N and *Post* : P x T $\rightarrow$ N are the backward and forward incidence mappings. These mapping are such that:

$\forall (P_i, T_j) \in P^C \times T^D$, Pre $(P_i, T_j)$ = Post $(P_i, T_j)$ = 0;

And : $\forall (P_i, T_j) \in P^D \times T^C$, Pre $(P_i, T_j)$ = Post $(P_i, T_j)$;

This means that no arcs connect C–places to D–transitions, and if an arc connects a D–place $P_i$ to a C–transition $T_j$, the arc connecting $T_j$ to $P_i$ must exist. This appears graphically as loops connecting D–places to C–transitions.

These two conditions mean that, in a D–elementary hybrid PN, only the discrete part may influences the continuous part behavior, the opposite never occurs (the continuous part has no influence on the discrete par).

4. $h: P \cup T \rightarrow \{C, D\}$ defines the set of continuous nodes, (h (x) = C) and discrete nodes, (h (x) = D).

5. $S: T^D \rightarrow R^+ \times (R^+ \cup \{\infty\})$ associates to each D–transition $T_j$ its firing interval $[\alpha_j, \beta_j]$.

6. $V: T^C \rightarrow R^+$ associates a maximal firing speed $V_j$ to each C – transition $T_j$.

7. $M_0$ is the initial marking; C–places contain non-negative real values, while D–places contain non-negative integer values.

❑

*Example 3:* Consider the system of tanks and suppose that valves 1 and valve 2 may be into the two states open and closed. The passage from the open state to the closed state takes 3 t.u, but the commutation decision can be delayed indefinitely for the design of a control for example. It is why the time interval [3, ∞] is associated with discrete transitions $T_2$ and $T_3$. On the other hand, the passage from the closed state to the open state takes place after 10 *t.u.* from the last opening action. It is why the time interval [10, 10] is associated with discrete transition $T_1$ and $T_4$. The D–elementary hybrid PN in Figure 4 describes this hybrid system.

As a D–elementary hybrid PN combines a discrete and a continuous PN, its state at time t is given by the states of the two models. The strong coupling of these models makes it complex to analyze the hybrid model. Translating it in a hybrid automaton (HA) permits the use of tools and techniques developed for HA analysis. In [8], the authors developed an algorithm permitting translation of D–elementary hybrid PN into a HA. In the sequel, we briefly present this algorithm.

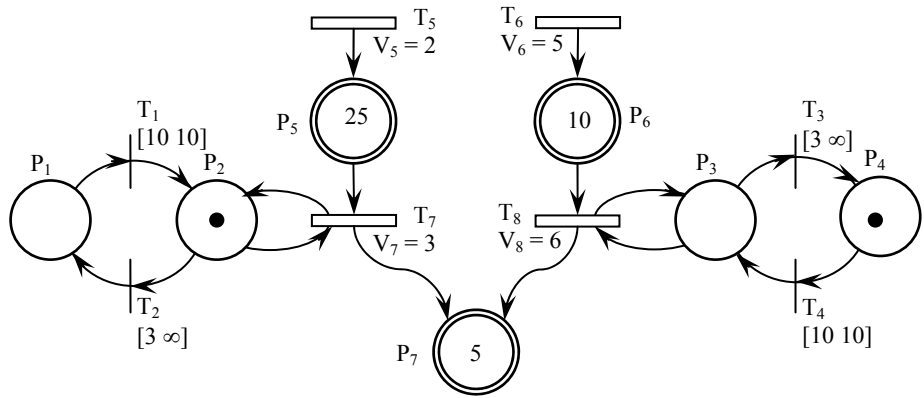

Figure 4. D–elementary hybrid Petri net describing the system of tanks

### III.2. Translating D – elementary hybrid PNs in hybrid automata

Hybrid automata were introduced by alur *et al*, [16] as an extension of finite automata, which associate a continuous dynamics to each location. It is the most general model in the sense that it can model the largest continuous dynamics variety. A HA is defined as follow

***Definition 4 (Hybrid Automata):*** An n – dimensional HA is a structure $\mathcal{HA} = (Q, X, L, T, F, Inv)$ such that:

1. $Q$ is a finite set of discrete locations;

2. $X \subseteq R^n$ is the continuous state space; it is a finite set of real-valued variables; A valuation $v$ for the variables is a function that assigns a real-value $v(x) \in R$ to each variable x $\in$ X; *V* denotes the set of valuations;

3. $L$ is a finite set of synchronization labels;

4. $\delta$ is a finite set of transitions; Each transition is a quintuple T = *(q, a, μ, γ, q')* such that:

   - $q \in Q$ is the source location;
   - $a \in L$ is a synchronization label associated to the transition;
   - $\mu$ is the transition guard, it is a predicate on variables values; a transition can be taken whenever its guard is satisfied;
   - $\gamma$ is a reset function that is applied when taking the corresponding transition;
   - $q' \in Q$ is the target location;

5. $F$ is a function that assigns to each location a continuous vector field on X; While in discrete location $q$, the evolution of the continuous variables by the differential equation $\dot{x} = f_q(x)$; This equation defines the dynamics of the location $q$;

6. *Inv* is a function that affects to each location $q$ a predicate *Inv* (*q*) that must be satisfied by the continuous variables in order to stay in the location $q$;

❑

A state of a HA is a pair (*q*, *v*) consisting of a location q and a valuation *v*.

Several problems, related to analysis of HA properties, could be expressed as a reachability problem. Note that this problem is generally undecidable unless strong restrictions are added to the basic model, to obtain special sub – classes of HA, [17]. The existence of computer tools allowing the analysis of the reachability problem for some classes of HA makes that the analysis of several hybrid systems formalisms is made after their translation in HA [18][19][20].

It is, generally, very complex to translate a hybrid PN in a hybrid automaton because of the strong coupling between discrete and continuous dynamics. D – elementary hybrid PN represents only a class of hybrid PNs, it permits modeling of a frequently met actual systems: *i.e.* the class of continuous flow systems controlled by a discrete event system. The translation algorithm consists in separating the discrete and the continuous parts. Then, the translation in an automaton is performed in a hierarchical way. The algorithm is based on three steps as described briefly below:

1. Isolate the discrete PN of the hybrid model and construct its equivalent timed automaton
2. Construct the hybrid automaton corresponding to each location of the timed automaton resulting from the previous step
3. Replace transitions between macro-locations by transitions between internal locations

*Example 4:* Consider the D – elementary HPN in Figure 4. Its discrete part is set again in Figure 5.a. The timed automata corresponding to this time PN is represented in Figure 5.b.

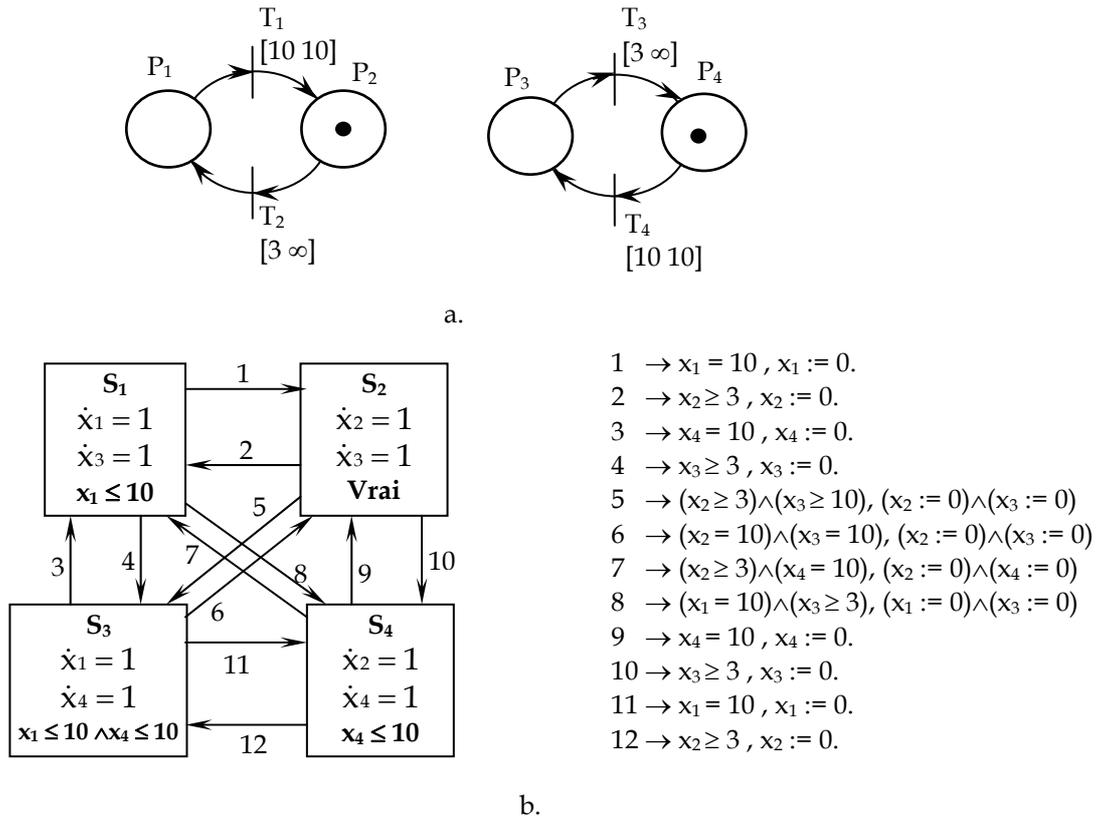

1 → $x_1 = 10$ , $x_1 := 0$.
2 → $x_2 \geq 3$ , $x_2 := 0$.
3 → $x_4 = 10$ , $x_4 := 0$.
4 → $x_3 \geq 3$ , $x_3 := 0$.
5 → $(x_2 \geq 3) \wedge (x_3 \geq 10)$, $(x_2 := 0) \wedge (x_3 := 0)$
6 → $(x_2 = 10) \wedge (x_3 = 10)$, $(x_2 := 0) \wedge (x_3 := 0)$
7 → $(x_2 \geq 3) \wedge (x_4 = 10)$, $(x_2 := 0) \wedge (x_4 := 0)$
8 → $(x_1 = 10) \wedge (x_3 \geq 3)$, $(x_1 := 0) \wedge (x_3 := 0)$
9 → $x_4 = 10$ , $x_4 := 0$.
10 → $x_3 \geq 3$ , $x_3 := 0$.
11 → $x_1 = 10$ , $x_1 := 0$.
12 → $x_2 \geq 3$ , $x_2 := 0$.

b.

Figure 5. Time PN and its equivalent timed automata

With each location of the timed automaton, corresponds a marking of the time PN, and therefore a configuration of the CCPN. For instance, if $P_2$ is unmarked, $T_7$ may be eliminated from the CCPN. The location $S_1$, for example, corresponds to the time PN marking vector $[m_1\ m_2\ m_3\ m_4]^T = [1\ 0\ 1\ 0]^T$, for which the continuous part is reduced to CCPN in figure 6.a. This CCPN may be translated in the HA in figure 6.b.

After the second step of the translation algorithm, we obtain a hierarchical form of a HA, formed from macro-locations containing each a HA describing the continuous dynamics in it. A generic representation of the model resulting after step 2 of the algorithm is given in figure 7.

The locations number of the resulting hybrid automaton depends on two parameters, i) locations number of the TA describing the discrete part behavior, denoted as *n*; ii) continuous places number of the continuous part, denoted as *m*. The first parameter *n* is finite for bounded time PN, although the propriety of boundedness is undecidable for time PN, it exists restrictive sufficient conditions for its verification [21]. This first parameter defines the macro – locations number. The second parameter m defines the number of locations inside a macro-location. As mentioned before, we can always model the behavior of a continuous PN by a HA with a finite number of locations, even if the continuous PN is unbounded, this number is least or equal to $2^m$. We have therefore a resulting HA that contains at the most $(n.2^m)$ locations. This is an important result since, it is generally impossible to bound a priori the number of reachable states in a hybrid PN.

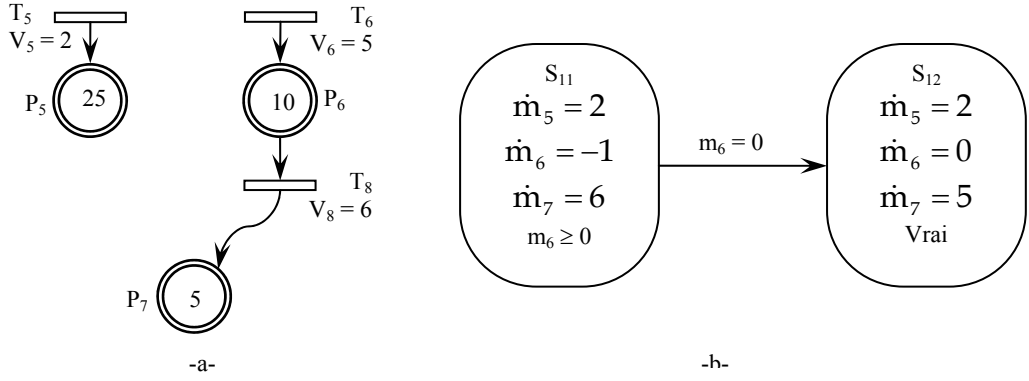

Figure 6. CCPN and its equivalent timed automata.

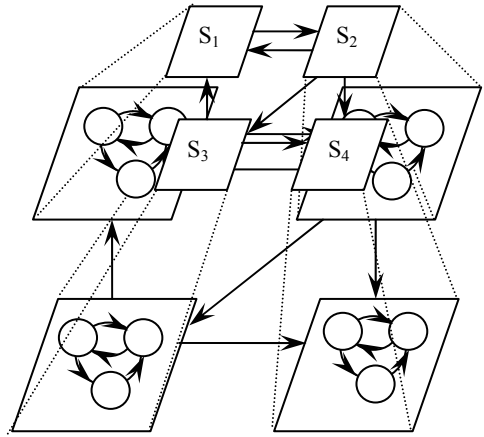

Figure 7. Generic schematization of model resulting from the second step of the algorithm.

### III.3. Hybrid Petri nets

A hybrid PN is distinguished from D–elementary hybrid PN in the fact that the first one contains a T–timed PN for modeling the discrete part. Timed fixed values are associated with each transition. Whereas, in the second model contains a T–time PN.

***Definition 5 (hybrid Petri Net) :*** A hybrid PN is a structure $\mathcal{PN}_H = (P, T, Pre, Post, h, S, V, M_0)$ such that :

1. $P = \{P_1, P_2, \ldots, P_n\}$ is a finite set of m places. $P = P^D \cup P^C$;

2. $T = \{T_1, T_2, \ldots, T_m\}$ is a finite set of n transitions. $T = T^D \cup T^C$;

3. $Pre$ : P x T $\rightarrow$ N and $Post$ : P x T $\rightarrow$ N are the backward and forward incidence mappings. These mapping are such that:

$\forall (P_i, T_j) \in P^D \times T^C$, Pre $(P_i, T_j)$ = Post $(P_i, T_j)$;

4. $h$ : P$\cup$T$\rightarrow$\{C, D\} defines the set of continuous nodes, (h (x) = C) and discrete nodes, (h (x) = D).

5. $S : T^D \rightarrow Q^+$ associates to each D – transition $T_j$ a duration $d_j$.

6. $V : T^C \rightarrow R^+$ associates a maximal firing speed $V_j$ to each C – transition $T_j$.

7. $M_0$ is the initial marking.

❑

The condition on backward and forward incidence mappings means that, if an arc connects a D–place $P_i$ to a C–transition $T_j$, the arc connecting $T_j$ to $P_i$ must exist. And vice versa. This appears graphically as loops connecting D–places to C–transitions. It means that a discrete token cannot be split by a continuous transition. The hybrid PN model, as defined below, allows modeling of the logical conditions, but it allows also the modeling of transformation of a continuous flow into discrete parts and vice versa.

*Example 5 :* Let us consider again the system of three tanks, and suppose that we have the following control strategy: we want to keep the liquid level in tank 1 more than 20 v.u., and in tank 2 more than 12 v.u. The hybrid PN in Figure 8 describes a system that satisfies this specification on tanks level. The weights 17 and 9, associated with the arcs correspond to the minimal thresholds of the two tanks taking into account the delay 3. Figure 9 illustrates the evolution of $P_5$ and $P_6$ marking in time.

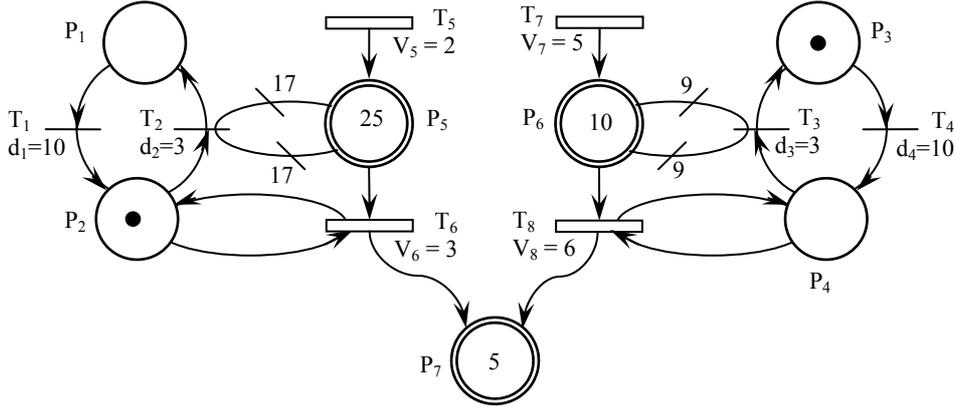

Figure 8. Hybrid Petri net describing the system of tanks with a restriction on its marking.

## IV. Controller synthesis

The controller Synthesis of HDSs drifts directly from Ramadge and Wonham theory [22]. Those, from a discrete event system synthesize a controller whose role is to forbid the occurrence of certain events. The controller decision to forbid an event depends only on the past of the system, *i.e.* of events, which already occurred. The aim is that the system coupled to its controller respects some given criteria.

Many researches were devoted to the problem of controller synthesis autonomous discrete event systems. This problem is thus well solved for this category of systems. The number of works relative to real – time systems controller synthesis is also very significant [23]. However, few works were devoted for solving this problem for HDS [24], [25], [26] and [27].

The controller synthesis of a dynamic system (autonomous, timed or hybrid) is generally based on three steps:

 i) the behavioral description of the system (called open loop system) by a model;
 ii) the definition of specifications required on this behavior;
 iii) the synthesis of the controller which restricts the model behavior to the required one, using a controller synthesis algorithm.

These algorithms consider the open system S and the specification on its behavior ϕ and try to synthesize the controller C, so that the parallel composition of S and C (S || C) satisfies ϕ. These algorithms use traditionally automata (finite state automata, timed automata and hybrid automata) because of their ease of formal manipulation; however, a model like HPN is preferred in the first step (the step of behavior description)

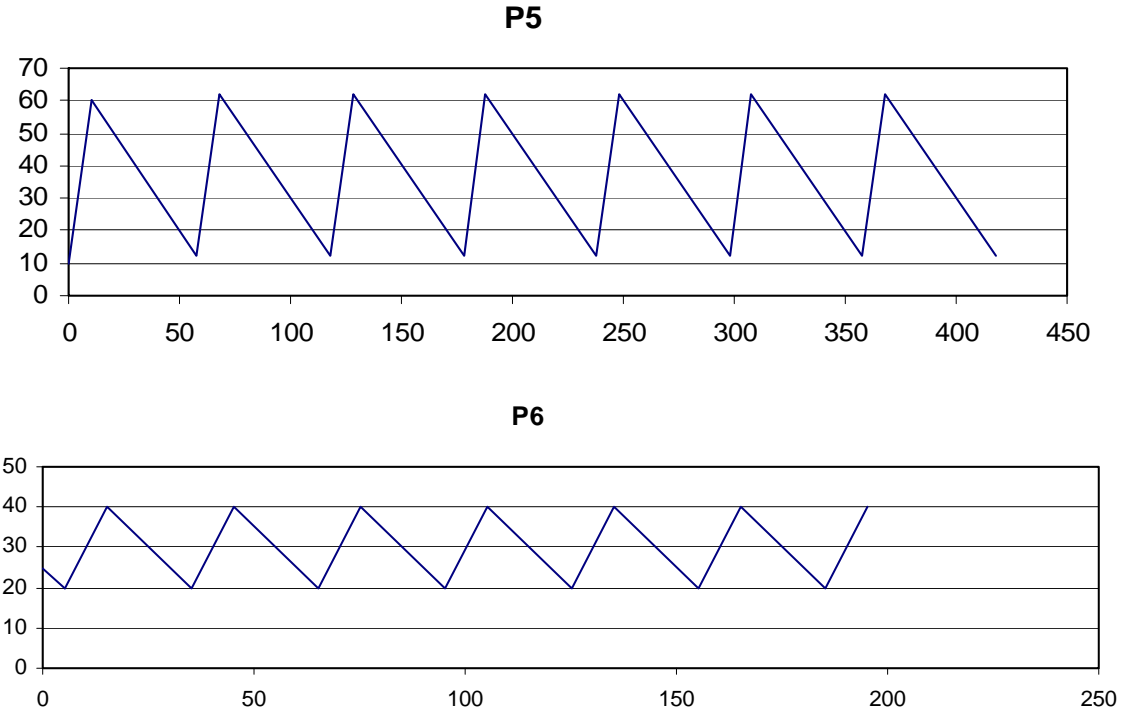

Figure 9. Temporal evolution of the marking of places $P_5$ and $P_6$ in HPN figure 9.

Consider an Open loop Hybrid system, the aim of controller synthesis is to construct a controller that satisfies the specifications closed loop hybrid system. These specifications imply, generally, restrictions on the closed loop hybrid system. They can be either (1) *specifications on the discrete part* (this type of specification forbid certain discrete states); or (2) *specification on the continuous part*, in this case the specification has the form of an invariant that the continuous state must satisfy. This implies that the continuous state of the closed loop hybrid system is restricted to a specified region. The open problem is to synthesize the guards associated with the controllable transitions so that the specifications are respected leading to a maximal permissive controller;

## V.     Conclusion

In this paper, we have presented some extension of PNs permitting HDSs modeling. The first models to be presented are continuous PNs. This model may be used for modeling either of a continuous system or of a discrete system. In this case, it is an approximation often satisfactory.

Hybrid PNs combine in the same formalism a discrete PN and a continuous PN. Two hybrid PNs model were considered in this paper. The first, called hybrid PN has a deterministic behavior; it means that we can predict the occurrence date of any possible event. The second hybrid PN considered is called D – elementary hybrid PN, this model was conceived to be used for HPNs controller synthesis.

Controller synthesis algorithms consider the open system S and the specification on its behavior $\phi$ and try to synthesize the controller C, so that the parallel composition of S and C (S || C) satisfies $\phi$. These algorithms use traditionally automata (finite state automata, timed automata and hybrid automata) because of their ease of formal manipulation; however, this model is not the most appropriate for behavior description. For coupling the analysis power of hybrid automata with the modeling power of hybrid PNs, an algorithm permitting translation of D – elementary hybrid PNs in hybrid automata was presented. Our future research is to generalize the existing results to the control of hybrid systems modeled by hybrid PNs.